\documentclass[12pt,epsf]{article}
\usepackage{graphicx}
\usepackage{eufrak}
\usepackage[mathscr]{eucal}
\usepackage{latexsym}
\usepackage{epsfig}

\def\be{\begin{equation}}
\def\ee{\end{equation}}
\def\ba{\begin{eqnarray}}
\def\ea{\end{eqnarray}}
\def\kk{\mathfrak{K}}

\def\lL{\mathscr{L}}
\def\hh{\mathscr{H}}

\begin{document}
\baselineskip=15.5pt
\pagestyle{plain}
\begin{titlepage}

\rightline{hep-th/0007064}

\rightline{HUTP-00/A028}

\begin{center}

\vskip 1.7 cm

{\LARGE {\bf Quantum Cosmology and $AdS/CFT$}}

\vskip 2.5cm
Luis Anchordoqui$^{a,}$\footnote{doqui@hepmail.physics.neu.edu},
Carlos Nu\~nez$^{b,}$\footnote{nunez@lorentz.harvard.edu}
and Kasper Olsen$^{b,}$\footnote{kolsen@feynman.harvard.edu}

\medskip
\medskip
${}^a${\it Department of Physics, Northeastern University\\
Boston, MA 02115, USA}

\medskip
${}^b$
{\it Department of Physics, Harvard University\\
Cambridge, MA  02138, USA}

\vspace{2cm}

\end{center}

\noindent
In this paper we study the creation of brane--worlds in $AdS$ bulk.
We first consider the simplest case of onebranes in $AdS_3$.
In this case we are able to
properly describe the creation of a spherically symmetric brane-world
deriving a general expression for its wavefunction. Then,
we sketch the $AdS_{d+1}$ set--up within the context of the WKB
approximation. Finally, we comment on these scenarios in light of
the $AdS/CFT$ correspondence.

\date{07/2000}

\end{titlepage}

\newpage

\tableofcontents

\bigskip

\section{Introduction}

The idea that spacetime has more than four dimensions is actually
quite old. Already in the 1920's, Kaluza suggested that gravity and
electromagnetism can be interpreted as the degrees of freedom of
the metric of a five-dimensional spacetime \cite{kk}. Later, Klein
\cite{klein} gave an explanation for the fact that the extra dimension
is not observed by suggesting that the extra dimension is compact and
very small.
Since then the idea has been studied from many different
perspectives {\it e.g.} in Kaluza-Klein supergravity theories and also
in string theories -- where more than three spatial dimensions naturally arise
but the extra dimensions are usually assumed to be of Planck size
for not been directly observable.
In another direction
it has been suggested \cite{RSV} that spacetime can
have more than three {\it noncompact} spatial dimensions if we live on a
four-dimensional domain wall which is embedded in the higher dimensions.
More recently, there has been a renewed interest in the topic
since progresses in string theories \cite{PHW} have modified the old scenario
(where the extra dimensions cannot exceed the tiny scale
$\sim 1$ TeV$^{-1} \sim 10^{-19}$ m) suggesting that Standard Model gauge
interactions could be confined to a four-dimensional subspace
-- or brane--world -- whereas gravity can still propagate in the whole bulk 
spacetime. Actually, the possibility that part of the standard model 
particles live in large (TeV) extra dimensions was first put forward in 
connection to the problem of supersymmetry breaking in string theory \cite{a}.
These scenarios
presents us with the enticing possibility to explain some
long-standing particle physics problems by geometrical means
\cite{arkani,rs,rs2}.

In the canonical example of \cite{arkani}, spacetime is a direct product of
ordinary four-dimensional spacetime and a (flat) spatial $d$-torus of
common linear size $r_c$. Within this simple model,
the large hierarchy between the weak scale and the fundamental
scale of gravity can be eliminated. However, the hierarchy only
arises in the presence of a large volume for the compactified
dimension which is very difficult to justify. A more compelling
scenario was introduced by Randall and Sundrum (herein RS).
Reviving an old idea \cite{RSV}, RS  proposed a set--up with the
shape of a gravitational condenser in which two branes of opposite tension
(which gravitationally repel each other) are stabilized by a slab of
anti-de Sitter (AdS) space \cite{rs}. In this model the extra
dimension is strongly curved, and the distance scales on the brane
with negative tension are exponentially smaller than those on the
positive tension brane. Such exponential suppression can then
naturally explain why the observed physical scales are so much
smaller than the Plank scale. In further work RS found that
gravitons can be localized on a brane which separates two patches
of $AdS_5$ spacetime \cite{rs2}, suggesting that it is possible
to have an infinite extra dimension \cite{lr}. The question
whether this scenario reproduces the usual four-dimensional gravity
beyond the Newton's law has been analyzed \cite{gravity} and
cosmological considerations of
models with large extra dimensions confirms that they are at least
consistent candidates for describing our world \cite{cosmology}. These
ideas have raised a lot of interest in the subject and several groups
have begun to work on possible experimental signatures of the extra
dimension(s) \cite{l3}.

In this paper we shall discuss the creation of brane--worlds in
$AdS$ bulk. The approximation scheme to be used is the
minisuperspace restriction of the canonical Wheeler--DeWitt
formalism.  The basic idea of this approach, commonly adopted in
quantum cosmology calculations \cite{mini}, is to separate the
space-like metric into ``modes'', and then insist that all the
``translational'' modes are ``frozen out'' by using the classical
field equations, leaving only the scale factor to be quantized.
The outline of the paper is as follows. We begin in section 2
by deriving a brane-big-bang in $AdS_3$. This lower--dimensional
model provides a simple setting in which certain basic physical
phenomena can be easily demonstrated while avoiding the
mathematical complexities  associated with the
higher--dimensional counterparts. In section 3 we
consider multi-dimensional brane-worlds, discussing the possible
cosmologies within the WKB approximation. In section 4 we analyze
the implications of the $AdS/CFT$ correspondence to quantum
cosmology.

\section{Brane--world in $AdS_3$}

\subsection{Wheeler--DeWitt Equation}

In this section we consider
the creation of a onebrane in $AdS_3$ within the framework of quantum
cosmology \cite{mini}. Thus the universe will initially
be described by three--dimensional Anti-de Sitter space in which 
onebrane bubbles can nucleate spontaneously.
As we shall see below, these bubbles appear (classically) 
at a critical size and then expand. 

We thus begin by considering the action for a onebrane coupled to
gravity,
\begin{equation}
S_{\rm tot}={L_p \over 16 \pi}\int_\Omega d^3x\sqrt{g} \left(R+ {2
\over \ell^2}\right)+   {L_p \over 8 \pi} \int_{\partial \Omega}
d^2x \sqrt{\gamma} \,\, \kk  + T \int_{\partial \Omega} d^2x
 \sqrt{\gamma},
\label{adsthreeaction}
\end{equation}
where $\kk$ stands for the trace of the extrinsic curvature of the
boundary, $\gamma$ is the induced metric on the brane, and $T$ is
the brane tension.\footnote{In our convention, the extrinsic
curvature is defined as $\kk_{\mu\nu} = 1/2 (\nabla_\mu\hat{n}_\nu
+ \nabla_\nu \hat{n}_\mu)$, where $\hat{n}^\nu$ is the outward
pointing normal vector to the boundary $\partial\Omega$. Lower Greek
subscripts run from 0 to 2, capital Greek subscripts from 0 to $d$, and
capital Latin subscripts from 0 to $(d-1)$. Throughout the paper we adopt
geometrodynamic units so that $G\equiv1$, $c\equiv1$ and $\hbar
\equiv L_p^2 \equiv M_p^2$, where $L_p$ and $M_p$ are the Planck
length and Planck mass, respectively.} The first term is the usual
Einstein-Hilbert (EH) action with a negative cosmological constant
($\Lambda=-{1 / \ell^2}$). The second term is the Gibbons-Hawking
(GH) boundary term, necessary for a well defined variational problem
\cite{gibbons-hawking}. The third term corresponds to a constant
``vacuum energy'', {\it i.e.} a cosmological term on the boundary.

We wish to consider a brane which bounds two regions
of $AdS_3$. If we further specialize to the case of spherical
symmetry \cite{deser-jackiw} where
\begin{equation}
 ds^2_3 = - \left(1 + \frac{y^2}{\ell^2} \right) dt^2 +
 \left(1 + \frac{y^2}{\ell^2} \right)^{-1} dy^2 +
 y^2 d\phi^2,
\label{threemetric}
\end{equation}
the geometry is uniquely specified by a single degree of freedom, the
``radius'' of the brane $A(\tau)$. The $\tau$ coordinate denotes
proper time as measured along the brane-world. The computation of the
GH boundary term has now reduced to
that of computing the two non-trivial components of the second
fundamental form. From Eq. (\ref{threemetric}) we find (see Appendix for
details):
\be
\kk_\phi^\phi =  \frac{1}{A} \left[1 + \frac{A^2}{\ell^2} +
\dot{A}^2 \right]^{1/2}, \ee and
\be
\kk_\tau^\tau =  \left[\ddot{A} + \frac{A}{\ell^2}\right] \,
\left[1+ \frac{A^2}{\ell^2} + \dot{A}^2 \right]^{-1/2},
\ee
(where the dot denote a derivative with respect to proper time).
After integration by parts the gravitational Lagrangian restricted
to this minisuperspace may be identified as,
\be \lL = \frac{L_p}{2} \left\{- \dot{A}\,\,{\rm arcsinh} \left[
\frac{\dot{A}}{\sqrt{1+A^2/\ell^2}} \right] +  \sqrt{1 +
\frac{A^2}{\ell^2} + \dot{A}^2} \right\}.
\ee
The classical Wheeler--DeWitt Hamiltonian is now easily extracted.
In order to do this we compute the conjugate momentum to $A$,
\be p = \frac{\partial \lL}{\partial
\dot{A}} = - \frac{L_p}{2} \,{\rm arcsinh} \left[
\frac{\dot{A}}{\sqrt{1+A^2/\ell^2}} \right].
\ee
This relation may be inverted to yield, $ \dot{A} = -  (1 +
A^2/\ell^2)^{1/2} \, \sinh (2 p/L_p)$, so that the Wheeler--DeWitt
Hamiltonian is
\be \hh_{\rm tot} \equiv p \dot{A} -
\lL_{\rm tot} =  - 2 \pi AT - \frac{L_p}{2} \sqrt{1
+\frac{A^2}{\ell^2}}\cosh (2p/L_p).
\label{machine}
\ee
Eq. (\ref{machine}) can be rewritten as,
\be
\hh_{\rm tot} = - 2\pi AT - \frac{L_p}{2} \,\sqrt{1+ \frac{A^2}{\ell^2} +
\dot{A}^2}.
\ee
The Hamiltonian constraint -- which follows from the requirement of
diffeomorphism invariance -- is
$\hh_{\rm tot} = 0$, or equivalently,
\be \dot{A}^2 = - 1
- A^2 \left( \frac{1}{\ell^2} - 16 \pi^2 \frac{T^2}{L_p^2} \right).
\label{EE}
\ee
Observe that the constraint equation is consistent with the
covariant conservation of the stress-energy tensor and reproduces
the classical Einstein field equations of motion. It is easy to
see from Eq. (\ref{EE}) that in order to obtain a real solution
we need $T \neq 0$. Furthermore, the brane--world is (classically)
bounded by a minimum radius
\be
A_0^2 = \left(\frac{ -1}{\ell^2} + \frac{16 \pi^2
T^2}{L_p^2}\right )^{-1},
\label{minrad}
\ee
with $16 \pi^2 \ell^2 T^2/L_p^2>1$. In other words, the brane bubbles
appear classically at a critical size and then their expansion is governed
by (\ref{EE}). Note that as the world
approaches the minimum size the expansion tends to zero. Once the world
is  dynamically stable it experiences
an everlasting expansion.
However, we shall soon see that quantum effects
permit well--behaved wave functions for vanishing $T$. With the
classical dynamics of the model understood and the Wheeler--DeWitt
Hamiltonian at hand, quantization is straightforward. Canonical
quantization proceeds via the usual replacement $p \rightarrow - i
\hbar
\partial/\partial A$. Naturally, the resulting quantum
Hamiltonian has a factor order ambiguity. This factor-ordering
ambiguity may be removed in a natural (though not unique) way by
demanding that the quantum Hamiltonian be Hermitian,
\be
\hat{\hh}_{\rm tot} = \frac{L_p}{2}\left(1 + \frac{A^2}{\ell^2}\right)^{1/4}
\cos \left[2 L_p \frac{\partial}{\partial A}\right] \left(1 +
\frac{A^2}{\ell^2}\right)^{1/4} + 2\pi AT. \label{w-dw}
\ee
That this Hamiltonian is Hermitian may formally be seen by
Taylor-series expansion of the cosine. A more precise statement is that
this Hamiltonian acts on the Hilbert space of square-integrable
functions defined on the half--interval $[0, \infty)$ subject to the constraint
$\psi(0) = 0$. This is most easily seen by noting that the
Hamiltonian is Hermitian on $L^2([0, \infty))$  only if
$\psi(0) =0$.\footnote{Here, the Hilbert space scalar-product is given by
the sum over all possible configurations (i.e. sizes)
of the brane-world.
Henceforth, for any operator $\hat{\cal O}$ acting over any
brane-wave-function
$\psi_j$, $<\psi_k|\hat{\cal O}|\psi_j> = \int_0^\infty
\psi_k^* \,\hat{\cal O}\,\psi_j\,dA$. In other words, $A$ is a parameter which
governs the evolution of the brane along the null geodesic congruence
of the extra dimension.}

The wave function of the brane-world is determined in the usual
fashion by the Wheeler--DeWitt equation $\hat{\hh}_{\rm tot} \psi(A) = 0$.
For the special case of $T =0$ we find the following
solution:
\be \psi_{mn} (A) = C_{mn} (\varphi_m -\varphi_n),
\ee
with
\be \varphi_j = \left(1+
\frac{A^2}{\ell^2}\right)^{-1/4} \exp\left[ -
\left(j+\frac{1}{2} \right) \frac{\pi}{2} \frac{A}{L_p} \right].
\ee
(See Fig.1. for a plot of some of these wavefunctions).
Here $m$ and $n$ are integer valued quantum numbers describing
the internal state of the brane. Negatives values of $m$, $n$ are not
normalizable and so need to be discarded, as is the case when $m=n$.
\begin{figure}[htb]
\epsfxsize=3.5in
\bigskip
\centerline{\epsffile{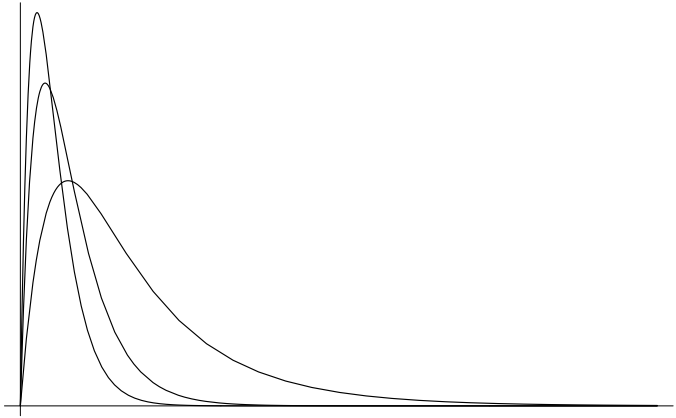}}
\caption{\baselineskip=12pt {\it Some sample wavefunctions for $T=0$ (upper is
$\psi_{2,3}$, middle is $\psi_{1,2}$ and lower is $\psi_{0,1}$).}}
\bigskip
\label{wavefunctionfig}\end{figure}
Note that the
appropriate normalization  is $\int|\psi|^2 dA =1$, and that
$\psi(0) = 0$, as required. In fact the two terms in $\psi_{mn}$
individually satisfy the differential equation $\hat{\hh}_{\rm
gravity} \psi =0$, but do not individually satisfy the boundary
condition. By appropriate choice of $C_{mn}$ these states may be
normalized, though they are not orthogonal to one another. The
normalization constant takes the rather complicated form:
\begin{eqnarray}
C_{mn} =\left\{ \frac{\ell\pi}{2} \left[
H_0(\ell(m+1/2)\pi/L_p)+H_0(\ell(n+1/2)\pi/L_p)-
2H_0(\ell(m+n+1)\pi/2L_p) \right. \right.\nonumber\\
\left. \left.  -  N_0(\ell(m+1/2)\pi/L_p)-N_0(\ell(n+1/2)\pi/L_p)+
2N_0(\ell(m+n+1)\pi/2L_p)
\right]\right\}^{-1/2},
\end{eqnarray}
where $H_0(z)$ is the Struve function and $N_0(z)$ is Neumann's
function\footnote{These functions have the following respective integral
representations:
$H_0(z)=\frac{2}{\pi}\int_0^{1}\frac{\sin(zt)\ dt}{\sqrt{1-t^2}}$ and
$N_0(z)=-\frac{2}{\pi}\int_1^{\infty}\frac{\cos(zt)\ dt}{\sqrt{t^2-1}}$.}.
With the wavefunctions at hand, one can calculate the mean value of the
``radius'' of the brane, {\it i.e.}
\begin{equation}
\langle A\rangle=\frac{\int A |\psi_{mn}|^2 dA}{\int|\psi_{mn}|^2 dA}.
\label{meanvalue}
\end{equation}
The integral in the numerator can be evaluated exactly but involves Meijer's
$G$-function $G^{31}_{13}$ and so the expression for $\langle A\rangle$ is
not of much practical use (anyhow, by dimensional analysis one would expect
that this number would be of order $L_p$). Using numerical integration we
have found $\langle A\rangle_{0,1}=0.54$, $\langle A\rangle_{1,2}=0.01$ and
$\langle A\rangle_{2,3}=10^{-3}$ in units of $L_p$ and for $\ell=1$.

\subsection{Qualitative Behaviour of Wavefunctions}

In this subsection we take a first look at the problem of finding solutions to
the Wheeler--DeWitt equation with $T\neq 0$ (in the next subsection we discuss
the solutions in the WKB approximation). The relevant equation is:
\be
\hat{\hh}_{\rm tot}\psi(A)=0
\ee
or, more explicitly
\be
\frac{L_p}{2}\left(1 + \frac{A^2}{\ell^2}\right)^{\frac{1}{4}}
\cos\left[2 L_p
\frac{\partial}{\partial A}\right]\left(1 + \frac{A^2}{\ell^2}
\right)^{1/4}\psi +
2 \pi TA \psi =0.
\ee
After defining $\varphi\equiv(1 + A^2/\ell^2)^{1/4}\,\psi$,
and the operator
\be
\Delta \equiv  \sum_{n=0}^\infty (-1)^n \frac{(2 L_p)^{2n}}{2
n!}\frac{\partial^{2n}}{\partial A^{2n}},
\label{eccompleta}
\ee
we see that we have to solve
\be
\Delta \varphi + \frac{4 \pi TA}{L_p\sqrt{1 +
A^2/\ell^2}} \, \varphi
=0.
\label{diffeq}
\ee
In order to gain some intuition for the behaviour of the solutions of this
equation
we will look for solutions in the two limits $A/\ell\gg 1$ and
$A/\ell\ll 1$. In the case $A/\ell\gg 1$, using a trial solution of the
form $\varphi = e^{\lambda A}$, we find that Eq. (\ref{diffeq}) reduces to
the following condition:
\be
\cos(2 L_p \lambda)= - 4\pi \frac{T\ell}{L_p}.
\ee
This is essentially the same condition as found before. Indeed,
if $16\pi^2\ell^2 T^2/L_p^2 > 1$ then $\lambda$ will have pure
imaginary values,
leading to an oscillatory solution at infinity, that is not acceptable
since it is not normalizable (it would in any case imply a delta function
normalization, that we are not considering here) and should be something
like a ``classical'' solution.  On the other hand, if
$16 \pi^2 \ell^2 T^2/L_p^2 \leq 1$,  $\lambda$ will have two real
solutions $\lambda_{\pm}$ of which one has to choose the
negative one, with the same criteria of normalizability as before.

\begin{figure}[htb]
\begin{center}
\epsfig{file=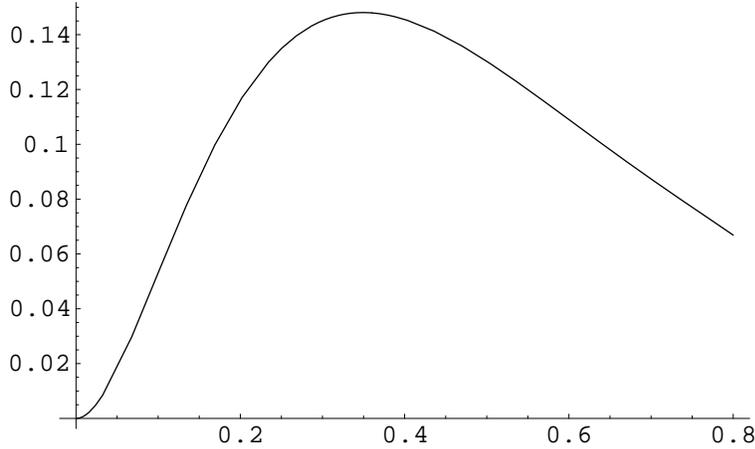,width=10.cm,clip=}
\caption{{\it Square of the wave function $\psi$ for small values of
the variable $a$ }}
\label{fig2a}
\end{center}
\end{figure}

In order to analyze more carefully the behaviour of the wave
function for $A/\ell \ll 1$, let us consider performing the
following change of variables \be A\to 2 L_p a. \ee With this
change of variables and after scaling $\psi\to\varphi$ as above,
the Wheeler--DeWitt equation reads: \be \tilde{\Delta} \varphi +
\frac{8 \pi T a }{\sqrt{1 + 4 L_p^2 a^2/\ell^2}}\varphi =0,
\label{diffeqs} \ee where \be
\tilde{\Delta}=\sum_{n=0}^\infty\frac{(-1)^n}{(2n)!}
\frac{\partial^{2n}}{\partial a^{2n}}. \ee It makes sense, since
the factor $L_p/\ell$ is small, to analyze the behaviour of this
equation for small values of the variable $a$ and expand the
square root in series.

The plot in Fig. 2 shows the behaviour, in the interval $[0,1]$, of the
square of the wave function $|\psi|^2$ that
solves numerically Eq.(\ref{diffeqs}),
where we have considered eighteen orders of derivatives in
$\tilde{\Delta}$. It is observed that $|\psi|^2$ has a maximum at 
$A\sim L_p$ as expected.
It is worth to point out that the solutions with less derivatives have
similar behaviour in the interval considered. It should be interesting to
find a method to analyze the complete series.

\begin{figure}[htb]
\begin{center}
\epsfig{file=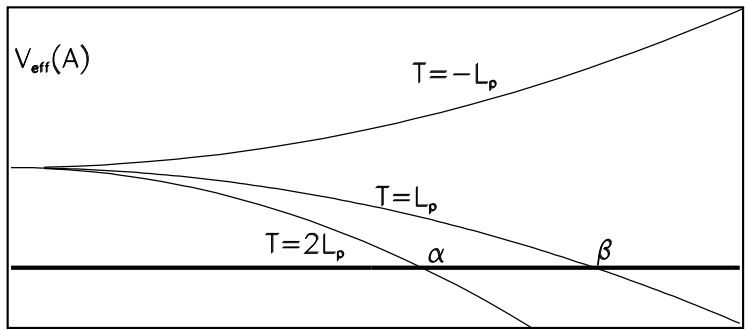,width=10.cm,clip=}
\caption{{\it Effective potential $V_{eff}(A)$
as a function of $A$ with the $AdS$ radius $\ell= 1$. $\alpha$ and $\beta$
are turning points.}}
\label{fig2}
\end{center}
\end{figure}

Let us now consider a qualitative analysis of the possible wavefunctions.
For this, let us set the $AdS$ radius to one, and analyze
the behaviour for
different relations between $T$ and $L_p$.
In Fig. 3 we plot a schematic representation of the potential energy
$V_{\rm eff} = A^2 ( 1/\ell^2 + 16\, \pi^2 \,T^2/L_p^2)$. Classically,
motion is confined to the region below the solid line (on which $V_{\rm
eff}=-1$). Strictly
speaking, when $T=2L_p$ classical motion is only
allowed for $A>\alpha$, while for $T=L_p$ the condition is
$A>\beta$. In this region the wave function $\psi$ presents
an oscillatory
behaviour modulated by $(1+A^2/\ell^2)^{-1/4}$, whereas from the
turning point to zero, $\psi$ is exponentially decreasing.
The complete shape of $\psi$ can be seen in  Fig. 4.
On the other hand, if $16 \, \pi^2 \ell^2 T^2/L_p^2< 1$, $V_{\rm eff}$
remains greater than $-1$ in the whole parameter space, and the classical
motion is always forbidden. In this case, $\psi$ can be expressed in terms
of exponentials with real arguments, yielding just vacuum fluctuations.
This is of course consistent with the behaviour we saw for $T=0$.

\begin{figure}[htb]
\begin{center}
\epsfig{file=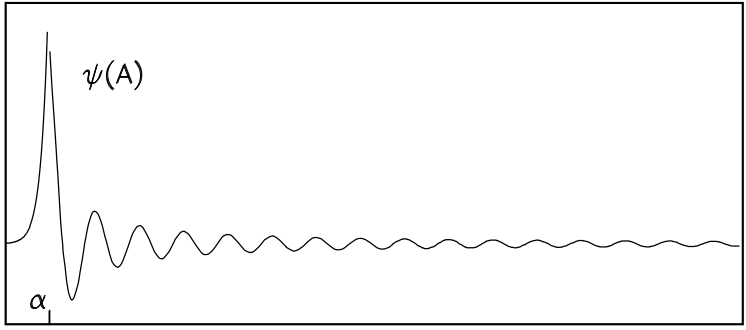,width=10.cm,clip=}
\caption{{\it Qualitative behaviour of the wavefunction. The turning point
is at $A=\alpha$.}}
\end{center}
\label{fig3}
\end{figure}

\subsection{WKB Approximation}

In this subsection we discuss solutions which are valid in the
near--classical domain. Since the potential is slowly varying
(see Fig. 3), one expects
the wave function to closely approximate the free particle state
wavefunction $\psi(A) = f(A) e^{i\,p\,A/\hbar}$.
Thus, we will look for solutions of the form
$\psi(A) = f(A) e^{iS(A)/\hbar}$. Following  \cite{visser}, the
semi-classical quantization condition may be written in the
generalized form,
\begin{equation}
\oint p(T,A) \, dA = (n_A+\delta)\hbar,
\end{equation}
where $n_A$ stands for the ``radial'' quantum number,
and $\delta$ is related
to the Maslov index \cite{brack}. For a Hamiltonian quadratic in
momenta, the usual WKB method shows that $\delta$ is typically a simple
fraction. In other cases, $\delta$
depends on both the Hamiltonian and the boundary conditions and is often
transcendental. In the present discussion, a precise
calculation of $\delta$ would add little to our understanding, thus, it
will not be evaluated but shall merely be carried along as an arbitrary
constant.

The precise form of the WKB wavefunction is determined by the following
constraint. In the semi--classical limit ($\hbar \rightarrow 0$),
the classical average in time of any quantity $Q(x)$,
\begin{equation}
\bar{Q}(x) = \frac{1}{\tau} \int_0^\tau Q(x(t)) dt = \frac{1}{\tau}
\int_0^\tau \frac{Q(x(t))}{v(x)} dx,
\end{equation}
has to be equal to the quantum average,
\begin{equation}
<\psi|Q|\psi> = \frac{\int |\psi(x)|^2 Q(x)dx}{\int |\psi(x)|^2 dx};
\end{equation}
where $v = \partial \hh /\partial p$, and the classical
time average $\tau = \int_0^\tau v^{-1} dx$.
Thus, the semi--classical approximation in the classically allowed region is
given by,
\begin{equation}
\psi_{{\rm WKB}}(A) =
\left|\frac{\partial \hh (p(T,A),A)}{\partial p}\right|^{-1/2} \,
\exp\left[ \pm \frac{i}{\hbar} \int^A p(T,x) dx \right],
\label{wkb1}
\end{equation}
while in the classical forbidden region it reads,
\begin{equation}
\psi_{{\rm WKB}}(A) =
\left|\frac{\partial \hh (p(T,A),A)}{\partial p}\right|^{-1/2} \,
\exp\left[ \pm \frac{1}{\hbar} \int^A p(T,x) dx \right].
\label{wkb2}
\end{equation}
It is easily seen that for the typical Hamiltonian quadratic in momentum
this generalized prescription reduces to the usual WKB approximation.

The conjugate momenta results in a multi-valued function:
\begin{equation}
p(T,A) = \pm \frac{L_p}{2}\left\{ {\rm arccosh} \left[
\frac{-4\pi TA}{L_p\sqrt{1+A^2/\ell^2}} \right] + 2 \pi i n \right \}.
\end{equation}
Here ${\rm arcosh} (x)$ is taken to map $[1,\infty)\rightarrow
[0,\infty)$, and $\pm$ refers to outgoing/ingoing directions.
In this scheme the imaginary contribution to
$p(T,A)$ does not contribute to the
quantization condition. The quantum number $n$, however, does
contribute when estimating the WKB wave function. In the classical
allowed region we get,
\begin{equation}
\psi_{{\rm WKB}}(A) =
\frac{\exp [- n\pi A/L_p]}
{|-1-A^2( 16 \pi^2 T^2/L_p^2 - 1/\ell^2)
|^{1/4}} e^{\pm i\Theta(A)},
\end{equation}
where
\begin{equation}
\Theta = \frac{1}{\hbar} \int^A \frac{L_p}{2}{\rm arccosh} \left[
\frac{- 4\pi Tx}{L_p\sqrt{1+x^2/\ell^2}} \right ] dx.
\label{integral}
\end{equation}
Note that in the limit $A/\ell <<1$,
\begin{equation}
\Theta = \frac{A}{2L_p} {\rm arccosh}\left[\frac{-4\pi TA}{L_p^2}\right].
\end{equation}
Thus, we recover the behaviour found in the previous subsection, $\psi$
exponentially increases from zero to the turning point.

If we now flip $T\rightarrow -T$,
and use ${\rm arcosh} (-x) = {\rm arccosh} (x) + i \pi$, we find that
\begin{equation}
\psi_{{\rm WKB}}(A) = \frac{\exp [- (n+1)\pi A/L_p]}
{|-1-A^2( 16 \pi^2 T^2/L_p^2 - 1/\ell^2)
|^{1/4}}e^{ \pm i\Theta(A)}
\end{equation}
are WKB eigenmodes corresponding to an eigenvacuumenergy $-T$.

This semiclassical solution blows up at
the turning points, where $\dot{A}$ goes to zero. This in itself may be
tolerated if the wavefunction is normalizable. The matching of the
wavefunction at the turning points may still be done by examining the wave
equation more closely in the vicinity of the turning point.

\section{Brane--world in $AdS_{d+1}$}

\subsection{Cosmology on the Brane}

We turn now to a more general analysis independent of the
dimension, i.e., for $AdS_{d+1}$ with $d>1$.
The expression for the total action is given by,
\begin{equation}
S_{\rm tot}= {L_p^{(3-d)} \over 16 \pi}\int_\Omega d^{d+1}x\sqrt{g}
\left(R+ {d\,(d-1)\over \ell^2}\right)+ {L_p^{(3-d)} \over 8 \pi}
\int_{\partial \Omega}
d^dx  \sqrt{\gamma} \,\, \kk  + T \int_{\partial \Omega} d^d x
 \sqrt{\gamma}.
\label{ogete}
\end{equation}
Let us also generalize the possible symmetries on the bulk which yield
different Robertson--Walker like cosmologies. The most general
$AdS_{d+1}$ metric can be written as,
\begin{equation}
ds^2 = -\left(k+\frac{y^2}{\ell^2}\right) dt^2 +
\left(k+\frac{y^2}{\ell^2}\right)^{-1} dy^2 + y^2 d\Sigma_k^2,
\end{equation}
where $k$ takes the values $0, -1, 1$ for flat, hyperbolic, or spherical
geometries respectively and where $d\Sigma_k^2$ is the corresponding metric
on the unit $(d-1)$-dimensional plane, hyperboloid, or sphere.
It should be stressed that if $k=-1$, an event horizon appears at
$y=\ell$. With this in mind, one can trivially
generalize the discussion in the appendix to get,
\be
\kk_{\phi_i}^{\phi_i} =  \frac{1}{A} \left[k + \frac{A^2}{\ell^2} +
\dot{A}^2 \right]^{1/2}, \ee and
\be
\kk_\tau^\tau =  \left[\ddot{A} + \frac{A}{\ell^2}\right] \,
\left[k+ \frac{A^2}{\ell^2} + \dot{A}^2 \right]^{-1/2},
\ee
where $i$ runs from 1 to $(d-1)$. In terms of these quantities,
the Einstein equation reads \cite{visser-book},
\begin{equation}
T g_{_{\Xi \Upsilon}} \delta^{^\Xi}_A \, \delta^{^\Upsilon}_B =
 \frac{L_p^{3-d}}{4\,\pi} [\kk_{AB} - {\rm tr} (\kk) g_{_{\Xi \Upsilon}}
\delta^{^{\Xi}}_A \, \delta^{^\Upsilon}_B].
\end{equation}
Its non--trivial components are,
\begin{equation}
T = - \frac{L_p^{(3-d)}}{4\pi} \frac{(d-1)}{A}\,\left(k+\frac{A^2}{\ell^2} +
\dot{A}^2 \right)^{1/2},
\label{m1}
\end{equation}
and
\begin{equation}
T = - \frac{L_p^{(3-d)}}{4\pi} \left\{
\frac{(d-2)}{A}\left(k+\frac{A^2}{\ell^2} +  \dot{A}^2 \right)^{1/2} +
\frac{\ddot{A} + A/\ell^2}{\sqrt{k +\dot{A}^2 + A^2/\ell^2}} \right\}.
\label{m2}
\end{equation}
It is easily seen that Eqs. (\ref{m1}) and (\ref{m2}) imply the conservation
of the stress energy. The evolution of the system is thus governed by,
\begin{equation}
\dot{A}^2 = -k - A^2 \left( \frac{1}{\ell^2} - \frac{16 \pi^2 \,T^2}{(d-1)^2\,
\,L_p^{2(3-d)}} \right).
\label{crazycosmology}
\end{equation}
A somewhat unusual feature of brane physics can be analyzed from
Eq. (\ref{crazycosmology}) (the five--dimensional case was already discussed
by Kraus, Ref. \cite{cosmology}).  Recall that in the spherical case,
the classical behaviour of the brane is bounded by a minimum
radius
\begin{equation}
A_0^2 = \left(-\frac{1}{\ell^2}
+\frac{16 \,\pi^2\,T^2}{(d-1)^2\,\,L_p^{2(3-d)}}\right)^{-1},
\end{equation}
but once the brane reaches that ``size'' it expands forever. Thus, contrary
to the standard Robertson Walker cosmology, the spherically symmetric brane
-- corresponding to $k=1$ --
represents an {\it open} world. Furthermore, depending on the
value of $T$ we can also  obtain a closed world with hyperbolic symmetry,
{\it i.e.} with $k=-1$.
On the one hand, if
\begin{equation}
\frac{16 \pi^2 T^2 \ell^2}{(d-1)^2 L_p^{2(3-d)}} \geq 1,
\label{geqone}
\end{equation}
the classical solution does not have turning points yielding an open world.
It should be remarked, however, that for $k=1$  the spacetime has no event
horizons, whereas if $k=-1$, the brane crosses an event horizon (at
$A = \ell$) in a finite proper time.

On the other hand, if
\begin{equation}
\frac{16 \pi^2 T^2 \ell^2}{(d-1)^2 L_p^{2(3-d)}} < 1,
\label{lessone}
\end{equation}
the classical solution has two turning points representing
a big--bang and a big--crunch. Again,
the spacetime has an event horizon at finite proper distance from the brane.
If $k=0$, one obtains a solution only if the inequality (\ref{geqone}) is
satisfied. In the critical, case the solution represents the
RS$_{d+1}$ brane--world. At this stage, it is noteworthy that a 
comprehensive analysis of a domain wall that inflates, either moving 
through the bulk or with the bulk inflating too, was first discussed by 
Chamblin--Reall \cite{cosmology}.

\subsection{Semiclassical Corrections}

With the field equations for an expanding $(d-1)$-brane in hand,
the generalization of the WKB approximation to $AdS_{d+1}$ is
straightforward. Of particular interest is $AdS_5$.\footnote{Note that
if $k=0$ and $T= 3/(4 \pi L_p \ell)$, one recovers the RS--world.}
Let us specialize again to the case of a spherically
symmetric brane.
In such a case, Eq. (\ref{ogete}) can be re--written as
\begin{eqnarray}
S_{\rm tot} & = & \frac{1}{L_p}\int d\tau
\left\{ -\frac{A^3}{3\ell^2}
\frac{\sqrt{\dot{A}^2 +
A^2/\ell^2 +1}}{1+A^2/\ell^2} + 3A \sqrt{1 + A^2/\ell^2 +
\dot{A}^2} \right. \nonumber \\
 & - & \left. 2 A\dot{A} \,{\rm arcsinh} \left[
\frac{\dot{A}}{\sqrt{1+A^2/\ell^2}} \right]
\right\} + T \int_{\partial \Omega}d^4x
 \sqrt{\gamma}.
\label{S-AdS_5}
\end{eqnarray}
For positive eigenvalues of $T$, the solution in the classical allowed
region is then given by,
\begin{equation}
\psi_{{\rm WKB}}(A) =
\frac{\exp [- 2 \,\pi\, n\, (A/L_p)^2]}
{|-1-A^2/\ell^2 + G^2)
|^{1/4}} \,e^{\pm i\, \int^A p\, dx},
\end{equation}
with $p\equiv\partial \lL_{5}/\partial\dot{A}$, and
$G(A) = 4 \pi A^2 T L_p/3$. The oscillating part will be a real
exponential term in the classically forbidden region.

\section{Relation to $AdS/CFT$ Correspondence}

\subsection{Generalities}

Another, seemingly different, but in fact closely related subject
we will discuss in this section is the $AdS/CFT$
correspondence \cite{malda}. This map provides a ``holographic'' projection
of the $AdS$ gravitational system into the physics of the gauge theory.
In the standard noncompact $AdS/CFT$ set up, gravity is decoupled from the
dual boundary theory. The prime example here
being the duality between Type IIB on
$AdS_5\times S^5$ and ${\cal N}=4$ supersymmetric $U(N)$
Yang-Mills in $d=4$ with
coupling $g_{YM}$ (the t'Hooft coupling is defined as
$\lambda=g_{YM}^2N$).
In this case it is known that the parameters of the $CFT$ are related to
those of the supergravity theory by \cite{malda,review}
\ba
\ell & = &\lambda^{1/4}l_s \\
\frac{\ell^3}{L_p^3}& = &\frac{2N^2}{\pi}, \label{string} \ea
where $l_s$ is the string length. The supergravity description is
valid when $\lambda$ and $N$ are large (so that stringy effects
are small). However, it is natural to suppose (in the spirit of
$AdS/CFT$) that any RS-like model should properly be viewed as a
coupling of gravity to whatever strongly coupled conformal theory
the $AdS$  geometry is dual to. In the following discussion, inspired 
in \cite{hawking}, we unfold on this hypothesis:
The most general action for a RS--like model in
$AdS_{d+1}$ is given by \be S_{RS}=S_{EH}+S_{GH}+2S_1+S_m, \ee
where $S_1$ is the counterterm $(T/2)\int d^dx\sqrt{\gamma}$. The
last term $S_m$ is the action for matter on the brane which was
not included in Eq. (\ref{ogete}), but it is included here for
completeness. Now, to apply the $AdS/CFT$-correspondence, there is
the question of the definition of the gravitational action in
$AdS_{d+1}$. The standard action -- corresponding to the two
first terms of Eq. (\ref{ogete}) -- is divergent for generic geometries
and one must add certain ``counterterms'' to obtain a finite
action \cite{BK,KLS}. Then we have schematically,

\be
S_{\rm grav}=S_{EH}+S_{GH}+S_1+S_2+S_3+\cdots ,
\ee
where $S_k$ is of order $2(k-1)$ in derivatives of the boundary
metric. Specifically, $S_2$ and $S_3$ are the counterterms 
discussed in \cite{hawking}.
They are expressed in terms of the boundary metric:
\be
S_2\,\propto\, \int d^dx\sqrt{\gamma} \tilde{R}
\ee
and
\be
S_3\, \propto\,
\int d^dx\sqrt{\gamma}\left(\tilde{R}_{ij}\tilde{R}^{ij}-\frac{d}{4(d-1)}\tilde{R}^2\right).
\ee
Some of the higher--order counterterms were computed in \cite{KLS}.
For a given dimension $d$, however, one only needs to add a {\it finite}
number of counterterms, specifically terms of order $2n < d$ in derivatives
of the boundary metric.

In \cite{BK} the counterterms were found for
$AdS_3$, $AdS_4$ and $AdS_5$ by requiring a finite mass density of the
spacetime. In the first case it was found, that
only $S_1$ is needed, while in the latter cases both $S_1$ and $S_2$ are
needed. Kraus et al. \cite{KLS} later derived a method for generating the
required counterterms for any dimension $d$.
Furthermore, in \cite{BK} it was also noted that for the case of $AdS_5$
one could add terms of higher order in derivatives of the metric, as for
example the counterterm $S_3$ but without changing the mass of the spacetime.
Confronted with this ambiguity we face the question of which counterterms
should be added in for example $AdS_5$.
For that, we note that in order to
apply the $AdS/CFT$--correspondence we should require that the symmetries on
both sides of the correspondence match. The Weyl anomaly was computed in
\cite{henningson} for gravity theories in $AdS_{d+1}$
and we can then apply this result to fix the
possible counterterms. For $d$ odd there is no such anomaly and the
divergent part of the (super)gravity action is canceled by the addition of
the above mentioned counterterms. This implies, for example, that for
$AdS_4$ we should only add $S_1$ and $S_2$. For $d$ even there is a
nonvanishing anomaly \cite{henningson}. For $AdS_3$
this means that both $S_1$ and $S_2$ should be added and for $AdS_5$ we
should add the terms $S_1$, $S_2$ and $S_3$. So, the requirement of 
finiteness of the action {\it
together} with the matching of Weyl anomalies fixes the precise form of the
supergravity action in $AdS_{d+1}$.

\subsection{Dual Boundary Theory}

Now, using the $AdS/CFT$-correspondence, one can easily
show that the RS--model in dimension $d+1$ is dual to a
$d$--dimensional $CFT$ (which we
call the RS $CFT$) with a coupling to
matter fields and the domain wall given by the action
$2S_2+2S_3+\cdots +S_m$, where we should remember that for $AdS_3$ 
and $AdS_4$, the $S_3$--term is
absent but appears in all higher--dimensional cases.
To illustrate this point,
let us now analyze the $AdS/CFT$ for the
simplest three--dimensional example. We will work in Euclidean space in
order to avoid definition problems in the path integral. In this case the
RS action (without matter) is given by
\begin{equation}
S_{RS}=-{L_p \over 16 \pi}\int_\Omega d^3x\sqrt{g} \left(R+ {2
\over \ell^2}\right)-  {L_p \over 8 \pi} \int_{\partial \Omega}
d^2x  \sqrt{\gamma} \,\, \kk - \frac{L_p}{4\pi}\int_{\partial \Omega} d^2x
 \sqrt{\gamma},
\label{adsthree}
\end{equation}
which is essentially the same as in Eq. (\ref{adsthreeaction}) but now with
the tension $T$ fixed to be $L_p/(4\pi)$. (More on this below).
Our set--up is as illustrated in Fig. 5: we have two regions $R_1$ and
$R_2$ bounded by a two--dimensional domain wall and on each of these
regions the metric is the $AdS_3$ metric $g_{ij}$ which induces the metric
$\gamma_{ij}$ on the wall.\footnote{For details of Penrose diagrams the
reader is referred to \cite{he}.} Following \cite{hawking,gubser}, let us
compute the partition function obtained by integrating over the bulk metric
with boundary value $\gamma_{ij}$ on the wall:
\be
Z_{RS}[\gamma] = e^{-2S_1}\left(\int_{R_1\cup R_2}{\cal D}g
\ e^{-S_{EH}[g]-S_{GH}[g]}\right),
\ee
where the integral is over the two patches $R_1$ and $R_2$ of $AdS$.
(Note that even though $S_{GH}$ is a two--dimensional term it depends on
the bulk metric through the extrinsic curvature of the domain wall and can
therefore not be taken out of the path integral).
Since the integral over the two regions of $AdS$--space are independent, we
can write it as an integral over a single patch of $AdS$--space\footnote{
Note that the result of the integral over the regions $R_1\cup R_2$ is
not the addition of the integrals, but the product. Indeed, since we are
dealing with independent processes, we have the product of the
probabilities
amplitudes instead of the sum, that would produce `interference effects'
not present in the RS set-up.}:
\begin{equation}
Z_{RS}[\gamma] = e^{-2S_1}\left(\int_{R_1}{\cal D}g\ e^{-S_{EH}[g]-
S_{GH}[g]}\right)^2.
\label{rspart}
\end{equation}

\begin{figure}[htb]
\begin{center}
\epsfig{file=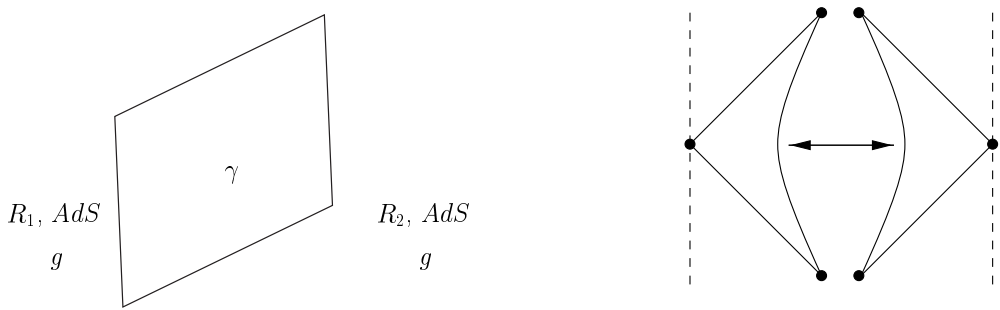,width=14.cm,clip=}
\caption{{\it Left: Schematic representation of two $AdS$ regions
bounded by a flat domain wall. Right: Penrose diagram of $AdS$ surgery.
The arrows denote identification and heavy dots represent points at infinity.
The dotted line denotes timelike infinity.}}
\end{center}
\label{fig007}
\end{figure}

Now according to the discussion above,
the partition function for a consistent gravity theory in $AdS_3$,
with finite mass of spacetime and appropriate central charge, is
\begin{eqnarray}
Z_{\rm grav}[\gamma] & = & \int_{[\gamma]}{\cal D}g\
e^{-S_{EH}[g]-S_{GH}[g]-S_1[\gamma] - S_2[\gamma]} \nonumber\\
 &  = &  e^{-S_1[\gamma]-S_2[\gamma]}
\int_{[\gamma]}{\cal D}g\
e^{-S_{EH}[g]-S_{GH}[g]} \nonumber \\
 & = & e^{-W_{CFT}[\gamma]},
\label{gravpart}
\end{eqnarray}
and according to the $AdS/CFT$ it should be identified with the generating
functional for connected Green's functions of the RS $CFT$ as above. By
combining Eq. (\ref{rspart}) and (\ref{gravpart}) we finally obtain:
\begin{equation}
Z_{RS}[\gamma] = e^{-2W_{CFT}[\gamma] + 2 S_2[\gamma]}.
\end{equation}
This shows that the RS--like model in $AdS_3$ is equivalent to a $CFT$
coupled to gravity with action $2S_2$. This dual gravity theory is actually
two--dimensional since $2S_2$ is the Einstein--Hilbert
action for two--dimensional gravity.
Similar correspondences can be
derived in higher--dimensional cases. For example we have:
\be
S_{RS}^{(4)}\leftrightarrow W_{RS}^{(4)}-2S_2+S_m,
\ee
while
\be
S_{RS}^{(5)}\leftrightarrow W_{RS}^{(5)}-2S_2-2S_3+S_m.
\ee
Here $W_{RS}$ stands for the generating functional of connected Green's functions
of the boundary (RS) $CFT$, that is twice the CFT induced on the brane.
Note that, as in the case of $AdS_3$, $-2S_2$ is the Einstein-Hilbert action for
$d$--dimensional gravity and so the RS model is equivalent to
$d$-dimensional gravity coupled to a $CFT$ with corrections to gravity
coming from the third counterterm $S_3$ (at least for $d>3$). 
This alone, however, does not tell us what the RS $CFT$ actually
is\footnote{The boundary $CFT$ can be found for the case of
$AdS_3$\cite{hms}.},
but rather that the RS model in $d+1$ dimensions can be viewed as a
$d$-dimensional gravity (including corrections) coupled to a $CFT$ with
matter.\footnote{Related ideas were discussed in \cite{no}.} 
And so, for example, in the case of $AdS_5$ this is another way to
see why gravity is trapped on the four--dimensional domain wall and why
there are corrections to Einstein gravity. (However,
there are no such corrections
in the case of $AdS_3$ and $AdS_4$ as we argued above).

\subsection{Physical Implications}

Up to this point we have kept the tension of the domain wall, $T$,
arbitrary. Because of the various bounds described in sections 2
and 3 for different behaviours of the braneworld, it is important
to see what one might expect. Let us again first restrict to
$AdS_3$ for simplicity. It is well known that gravity in
asymptotically $AdS_3$ spacetime has a holographic description as
a 1+1 dimensional conformal field theory with central charge $c=
3\, \ell\, M_p/2$ \cite{b-h}. In order to recover the geometry
discussed in section 2, one must glue two copies of such bounded
$AdS_3$ spacetimes, and then integrate over boundary metrics.
Consequently, one has two copies of the matter action on the
boundary, with total central charge $c= 3\,\ell\,M_p$. In
addition, if $\tilde{R}> 0$ the conformal anomaly of the $CFT$
increases the effective tension on the domain wall, $T> L_p /
4\pi\ell$, yielding a de Sitter universe with an effective
cosmological constant driving inflation.\footnote{A few words of 
caution; it is quite 
possible that the truncation from an infinite number of degrees of 
freedom down to only one degree of freedom, $A(\tau)$, has also drastically 
truncated the real physics. Unfortunately, a treatment using Wheeler's full 
superspace is beyond the scope of our present calculation abilities.}
An (early) inflationary epoch looks very promising. 
The tremendous expansion during inflation may blow up a small sized region 
of the world (which was causally connected before inflation) to a size much 
greater than our current horizon. Therefore, it can be expected that the 
observable part of the brane looks smooth and flat, regardless of the 
initial curvature of the brane that inflated.\footnote{Note that a flat 
Robertson-Walker Universe requires a total energy density equal to the 
critical density $\rho_{\rm cr}$, whereas ordinary matter contributes 
only about a 5\% of $\rho_{\rm cr}$. A novel solution to this problem 
consistent with a large body of observations is the so-called 
``Manifold Universe'' \cite{manifold-universe}.} Furthermore, if we
consider conformal matter on the brane the inflationary 
phase is unstable and could decay into a matter dominated universe 
with thermalized regions, in agreement with current observations \cite{staro}. 

Another interesting process which could lead to brane--world 
reheating is as follows: During inflation 
trapped regions of false vacuum (within their
Schwarzchild radii) caught between bubbles of true vacuum may give rise 
to the creation of primordial black strings.  
Now, it is well--known that the black string solution suffers from a 
Gregory--Laflamme instability \cite{gl} leading to the 
formation of stable black cigars on the brane.
In addition, it was shown in \cite{gs} 
that the nucleation of supermassive bulk black
holes is highly supressed compared to the above mentioned process.  
Thus, prompted by the conventional arena \cite{ca}, one could speculate that 
the Hawking--evaporation of primordial 
black cigars slows down inflation. 
On the other hand, one could assume the existence of such a bulk black hole. 
Even in this case, the (brane-world/bulk-black-hole) 
system evolves towards a configuration of 
thermal equilibrium as was recently shown in \cite{CKN}.

Let us now briefly discuss a general $n$-dimensional brane-world
that falls under the action of a higher dimensional gravitational field.
The system can be decomposed into falling shells (which do not interact
with each other or with the environment that generates the metric),
with trajectories described by the scale factor $A(\tau)$.
From the above discussion it is clear that the value of $T$ will depend on
the symmetries of the domain wall. It is easily seen, for instance,
that if $k=-1$ then
\be
T<\frac{(d-1)L_p^{3-d}}{4\pi \ell},
\label{adstension}
\end{equation}
yielding a closed universe. Roughly speaking, the cosmological
constant induced by the conformal anomaly accelerates/slows down the
brane to balance the null geodesic congruence in the bulk,
shirking the world's pinch off. We recall that if $k=-1$, the
spacetime has an undesirable event horizon that must be reached
by the brane in a finite proper time.

Despite the fact that it is contrary to the spirit of RS-worlds, it
would be nice to add ``matter fields'' in the bulk to study the quantum
cosmology and the dual $CFT$ coupled to gravity that in this case should be
deformed by the insertion of operators.

Even though many kind of interesting phenomena are recognized,
brane-world cosmology remains thoroughly non-understood. The lower
dimensional model  here discussed can hopefully illuminate
the ``physical $AdS_5$ cosmology''.

\subsection*{Acknowledgements}
We would like to thank Vijay Balasubramanian, Thomas Hertog, Finn Larsen,
Juan Maldacena, Francisco Villaverde, Matt Visser, and Allan Widom for 
useful discussions/correspondence. Special thanks go to Harvey Reall
for a critical reading of the manuscript and valuable comments. 
The work of LA and CN 
was supported by CONICET Argentina, and that of KO by the Danish Natural 
Science Research Council.

\newpage

\appendix
\section{Appendix}
\setcounter{equation}{0}

Here we present a calculation of the second fundamental form of the
metric in Eq. (\ref{threemetric}) (it should be remarked that
this calculation is a direct analog to that of Ref. \cite{matt},
and it is included just for the sake of completeness).

Let us start by introducing a Gaussian normal coordinate system in
the neighborhood of the brane.
We shall denote the one--dimensional surface swept out by the brane by
$\Sigma$. Let us introduce a coordinate system $\phi_\perp$ on
$\Sigma$. Next we consider all the geodesics which are orthogonal
to $\Sigma$, and choose a neighborhood $N$ around $\Sigma$ so that
any point $p \in N$ lies on one, and only one, geodesic. The first
coordinate of $p$ is determined by the intersection of this
geodesic with $\Sigma$. The full set of spatial coordinates is
then given by $(\phi_\perp;\eta)$, while the surface $\Sigma$
under consideration is taken to be located at $\eta=0$ so that Eq.
(\ref{threemetric}) can be rewritten as
\be ds^2 =  - \left(1 +
\frac{y^2}{\ell^2} \right) dt^2 + d\eta^2 + y^2 d\phi^2,
\ee
fixed by the relation, $dy/d\eta = (1 + y^2/\ell^2)^{1/2}$. The second
fundamental form in such a coordinate--system  reads
\be \left. \kk_{\mu\nu}
\equiv  \frac{1}{2} \frac{\partial g_{\mu\nu}}{\partial \eta}
\right|_{\eta = 0, y=A}, \label{sff}
\ee
and its non-trivial components are
\be \kk_t^t = \frac{A}{\ell^2} \left(1 +
\frac{A^2}{\ell^2}\right)^{-1/2}, \ee  \be \kk_\phi^\phi =
\frac{1}{A} \left(1 + \frac{A^2}{\ell^2}\right)^{1/2}.
\ee

To analyze the dynamics of the system, we permit the radius of the
brane to become a function of time $A \rightarrow A(\tau)$.
Recall that the symbol $\tau$ is used to denote proper time as
measured by co--moving observers on the brane--world. Let the
position of the brane be described by $x^\mu(\tau,\phi) \equiv
(t(\tau),A(\tau),\phi)$, so that the velocity of a piece of
stress-energy at the brane $(u^\mu u_\mu = -1)$ is
\begin{equation}
u^\mu \equiv \frac{dx^\mu}{d\tau} =
\left(\frac{dt}{d\tau}, \frac{dA}{d\tau},0 \right).
\ee
We remind the reader that
\be ds^2 =
-\left(1+\frac{A^2}{\ell^2}\right) dt^2 +
\left(\frac{dA}{dt}\right)^2 \left(1+ \frac{
A^2}{\ell^2}\right)^{-1}  dt^2 + A^2 d\phi^2
\end{equation}
so,
\be d\tau^2 = - dt^2 \left\{ -\left(1+\frac{A^2}{\ell^2}
\right) + \left(\frac{dA}{dt}\right)^2
\left(1+\frac{A^2}{\ell^2}\right)^{-1} \right\}
\ee
or equivalently,
\be d\tau^2 = - dt^2 \left\{
-\left(1+\frac{A^2}{\ell^2}\right)^2 +
\left(\frac{dA}{dt}\right)^2\right\} \left(1+
\frac{A^2}{\ell^2} \right)^{-1}.
\ee
Since
\be \frac{dA}{dt}
= \frac{dA}{d\tau} \frac{d\tau}{dt},
\ee
we first get,
\be
-\left(1+\frac{A^2}{\ell^2}\right)
\left(\frac{d\tau}{dt}\right)^2 = - \left( 1 +
\frac{A^2}{\ell^2}\right)^2 + \dot{A}^2
\left(\frac{d\tau}{dt} \right)^2
\ee
and then,
\be
\frac{dt}{d\tau} = \frac{\sqrt{\dot{A}^2 + A^2/\ell^2 +1 }}
{1+A^2/\ell^2}.
\ee
Let us denote by $\hat{n}^\mu$ the unit normal vector to the brane,
which satisfies $u^\mu \hat{n}_\mu =
0$ and $\hat{n}^\mu \hat{n}_\mu =1$; its components are
$ \hat{n}^\mu =
(\dot{A}/(1+A^2/\ell^2), (1 + A^2/\ell^2 +
\dot{A}^2)^{1/2}, 0)$, such that the coordinate $y$ is
increasing in the direction $\hat{n}^\mu$.
Thus we obtain
\be \left.
\kk_\phi^\phi = \frac{1}{y} \frac{\partial y}{\partial
\eta}\right|_{y=A} = \frac{1}{A} \left(1 +
\frac{A^2}{\ell^2} + \dot{A}^2\right)^{1/2}.
\ee
To evaluate $\kk_\tau^\tau$ one can proceed in two alternative ways.
First one can simply use the definition
$\kk_{\mu\nu}=\frac{1}{2}\nabla_{(\mu}\hat{n}_{\nu)}$, giving:
\be
\kk_{tt} = \frac{1}{2} \nabla_{(t} \hat{n}_{t)} = \frac{d\hat{n}_t}{d\tau}
\frac{d\tau}{dt} - \Gamma_{tt}^\eta \,\, \hat{n}_\eta = - \frac{1 + A^2/\ell^2}
{\sqrt{1+\dot{A}^2 + A^2/\ell^2}} (\ddot{A} + A/\ell^2),
\ee
that using
\be
\kk_{\tau\tau} = \frac{\partial x^\mu}{\partial x^\tau}\,\frac{\partial
x^\nu}
{\partial x^\tau} \kk_{\mu\nu},
\ee
immediately yields
\be
\kk_\tau^\tau = \kk_t^t = \frac{\ddot{A} + A/\ell^2}
{\sqrt{1+\dot{A}^2 + A^2/\ell^2}}.
\label{ktautau}
\ee
Alternatively, one can easily check this last result by observing that
\be
\kk_\tau^\tau
\equiv - \kk_{\tau\tau} = - u^\mu u^\nu  \kk_{\mu\nu} = -u^\mu
u^\nu \nabla_\mu \hat{n}_\nu = u^\mu \hat{n}_\nu \nabla_\mu u^\nu
= \hat{n}_\mu ( u^\nu  \nabla_\nu u^\mu) = \hat{n}_\mu q^\mu,
\ee
where $q^\mu$ is the four acceleration of the brane. Now, by the
spherical symmetry of the problem the four acceleration is
proportional to the unit normal, $q^\mu \equiv q \,\hat{n}^\mu$, so
$\kk_\tau^\tau = q$. To explicitly evaluate the four
acceleration, utilize the fact that $\xi^\mu \equiv
\partial_t^\mu \equiv (1,0,0)$ is a Killing vector for the underlying geometry.
At the brane, the components of this vector are
$\xi_\mu = (-[1+A^2/\ell^2],0,0)$, so that
$\xi_\mu \hat{n}^\mu = - \dot{A}$ and $\xi_\mu u^\mu = - (1 +
A^2/\ell^2 + \dot{A}^2)^{1/2}$. With this in mind, comparing
\be \frac{d}{d\tau} (\xi_\mu u^\mu) =
\xi_\mu\, q\, \hat{n}^\mu = - q\, \dot{A}, \ee
and
\be \frac{d}{d\tau} (\xi_\mu u^\mu) = - \dot{A}
\frac{A/\ell^2 + \ddot{A}}{\sqrt{1+A^2/\ell^2 +
\dot{A}}},
\ee
we get
\be
\kk_\tau^\tau =  \frac{A/\ell^2 +
\ddot{A}}{\sqrt{1+A^2/\ell^2 + \dot{A}}} =
\frac{d}{d\tau} \left\{ {\rm arcsinh} \left[
\frac{\dot{A}}{\sqrt{1 + A^2/\ell^2}} \right]\right\} +
\frac{A}{\ell^2} \frac{dt}{d\tau};
\ee
this result agrees with that of Eq. (\ref{ktautau}). Having calculated the
nontrivial components of the second fundamental form we can now derive a
simpler expression for the relevant gravity--action (\ref{adsthreeaction})
in $AdS_3$.
Since $\sqrt{g}\, d^3x \rightarrow 2\pi
\, A \,\,dA \,dt$ and $\sqrt{\gamma} \,d^2x \rightarrow 2
\pi\,A\,d\tau$ an integration by parts finally leads to
\be
S_{\rm gravity} = \frac{L_p}{2} \int d\tau \left\{ - \dot{A}\,\,
{\rm arcsinh}\left[ \frac{\dot{A}}{\sqrt{1 + A^2/\ell^2}}
\right] + \sqrt{1 +\frac{A^2}{\ell^2} + \dot{A^2}} \right\}.
\ee


\newpage

\begin{thebibliography}{99}

\bibitem{kk} T. Kaluza, Akad. Wiss. Phys. Math. {\bf K1}, 966
(1921).

\bibitem{klein} O. Klein, Z.Phys. {\bf 37}, 895 (1926).

\bibitem{RSV} V. Rubakov and M. Shaposhnikov, Phys. Lett. {\bf 125B}, 136
(1983); M. Visser, Phys. Lett. {\bf B159}, 22 (1985).

\bibitem{PHW} J. Polchinski, Phys. Rev. Lett {\bf 75}, 4724 (1995)
[hep-th/9510017]; P. Ho\v rava and E. Witten, Nucl. Phys. {\bf
B460}, 506 (1996) [hep-th 9510209], Nucl. Phys. B {\bf 475}, 94
(1996) [hep-th/9603142].

\bibitem{a} I. Antoniadis, Phys. Lett. B {\bf 246} (1990) 377. 

\bibitem{arkani}  N. Arkani-Hamed, S. Dimopoulos and G. Dvali, Phys.
Lett. B {\bf 429}, 263 (1998); I. Antoniadis, N. Arkani-Hamed,
S. Dimopoulos and G. Dvali, Phys. Lett. B {\bf 436}, 257 (1998).

\bibitem{rs} L. Randall and R. Sundrum,
 Phys. Rev. Lett. {\bf 83}, 3370 (1999) [hep-ph/9905221].

\bibitem{rs2}L. Randall and R. Sundrum, Phys. Rev. Lett. {\bf 83}, 4690
(1999) [hep-th/9906064].

\bibitem{lr} J. Lykken and L. Randall, JHEP {\bf 0006}, 014 (2000)
[hep-th/9908076].

\bibitem{gravity}
A. Chamblin and G. W. Gibbons, Phys. Rev. Lett. {\bf 84}, 1090 (2000)
[hep-th/9909130]; A. Chamblin,
S. W. Hawking and H. S. Reall, Phys. Rev D {\bf 61}, 065007 (2000)
[hep-th/9909205]; R. Emparan, G.
T. Horowitz and R. C. Myers, JHEP {\bf 0001}, 007 (2000)
[hep-th/9911043]; R. Emparan, G.
T. Horowitz and R. C. Myers, JHEP {\bf 0001}, 021 (2000)
[hep-th/9912135]; J. Garriga and T. Tanaka, Phys. Rev. Lett.
{\bf 84} 2778 (2000) [hep-th/9911055];
M. Sasaki, T. Shiromizu and K. Maeda,
[hep-th/9912233]; A Chamblin, C. Cs\'aki, J. Erlich and T. J.
Hollowood, Phys. Rev. D (to be published) [hep-th/0002076];
S. B. Giddings, E. Katz and L.
Randall, JHEP {\bf 0003}, 023 (2000) [hep-th/0002091];
C. Grojean, [hep-th/0002130].

\bibitem{cosmology}

H. A. Chamblin and H. S. Reall, Nucl. Phys. B {\bf 562}, 
133 (1999) [hep-th/9903225]; 
N. Arkani-Hamed, S. Dimopoulos, N. Kaloper,
and J. March Russell, Nucl. Phys. B {\bf 567}, 189 (2000)
[hep-ph/9903224]; 
N. Kaloper,  Phys. Rev. D {\bf 60}, 123506 (1999)
[hep-th/9905210]; C. Cs\'aki, M. Graesser, C. Kolda, J. Terning,
Phys. Lett. B {\bf 462}, 34 (1999) [hep-ph/9906513];
J. M. Cline, C. Grojean and G. Servant, Phys. Rev. Lett. {\bf 83},
4245 (1999) [hep-ph/9906523]; H. B. Kim
and H. D. Kim,  Phys. Rev. D {\bf 61}, 064003
(2000) [hep-th/9909053]; P. Kanti, I. I. Kogan, K. A. Olive and
M. Pospelov,  Phys. Lett. B
{\bf 468}, 31 (1999) [hep-ph/9909481];
J. Cline, C. Grojean and G. Servant, Phys. Lett. B
{\bf 472}, 302 (2000) [hep-ph/9909496]; P. Kraus,  JHEP {\bf 9912} 011 (1999)
[hep-th/9910149]; S. Nam, [hep-th/9911237];  C. Cs\'aki, M. Graesser, 
L. Randall and J.
Terning, [hep-ph/9911406]; D. Ida,  [gr-qc/9912002];
N. Kaloper,  Phys. Lett. B {\bf 474}, 269 (2000)
[hep-th/9912125]; M. Cveti\v c and J.
Wang, [hep-th/9912187]; S. Mukohyama, T. Shiromizu,
and K. Maeda, [hep-th/9912287];
L. A. Anchordoqui and S. E. Perez Bergliaffa, Phys. Rev. D (to be published)
[gr-qc/0001019];
K. Koyama and J. Soda, Phys. Lett. B (to be published) [gr-qc/0001033];
C. Cs\'aki, J. Erlich, T. J. Hollowood and J. Terning [hep-th/0003076];
N. Deruelle and T. Dole\v zel,  [gr-qc/0004021]; C. Barcel\'o and M.
Visser,  [hep-th/0004022]; C. Barcel\'o and M. Visser,
Phys. Lett. B {\bf 482}, 183 (2000) [hep-th/0004056];
H. Stoica, S. H. Henry Tye and I. Wasserman,
[hep-th/0004126]; R. Maartens,
[hep-th/0004166]; P. F. Gonzalez-Diaz,
[gr-qc/0004078]; D. Langlois, [hep-th/0005025]; V. Barger, T. Han, T. Li,
J. D. Lykken and D. Marfatia, [hep-ph/0006275].

\bibitem{l3} G. F. Giudice, R. Rattazzi and J. D. Wells, Nucl. Phys. B {\bf
544}, 3 (1999) [hep-ph/9811291];
M. Acciarri et al. (L3 Collaboration), Phys. Lett. B {\bf 470},
281 (1999) [hep-ex/9910056];
S. Cullen, M. Perelstein and M. E. Peskin, [hep-ph/0001166];
Z. K. Silagadze, [hep-ph/0002255];
C. Adloff et al. (H1 Collaboration),
[hep-exp/0003002]; K. Cheung, [hep-ph/0003306].

\bibitem{mini} J. B. Hartle and S. W. Hawking, Phys. Rev. D {\bf 28},
2960 (1983).

\bibitem{deser-jackiw} S. Deser and R. Jackiw, Ann.
Phys. {\bf 153}, 405 (1984).

\bibitem{gibbons-hawking} G. W. Gibbons and S. W. Hawking,
Phys. Rev. D {\bf 15}, 2752 (1977).

\bibitem{visser} M. Visser, Phys. Rev. D {\bf 43}, 402 (1991).

\bibitem{brack} M. Brack and R. K. Bhaduri, {\it Semiclassical Physics},
(Addison-Wesley, 1997).

\bibitem{visser-book} M. Visser, {\it Lorentzian Wormholes}, (AIP Press,
Woodbury, N.Y. 1995); see in particular, chapter 14.

\bibitem{malda} J. Maldacena, Adv. Theor. Math. Phys. {\bf 2}, 231
(1998) [hep-th/9711200].

\bibitem{review} O. Aharony, S. S. Gubser, J.
Maldacena, H. Ooguri, and Y. Oz, [hep-th/9905111].


\bibitem{BK} V. Balasubramanian and P. Kraus, Commun. Math. Phys. {\bf 208},
413 (1999) [hep-th/9902121].

\bibitem{KLS} P. Kraus, F. Larsen and R. Siebelink,
[hep-th/9906127].

\bibitem{hawking} S. W. Hawking, T. Hertog, H. S. Reall, Phys. Rev. D {\bf 62},
043501, (2000) [hep-th/0003052].


\bibitem{henningson} M. Henningson and K. Skenderis, JHEP {\bf 9807}, 023
(1998) [hep-th/9806087].


\bibitem{he} S. W. Hawking and G. F. R. Ellis,
{\it The Large Scale Structure of Spacetime}, (Cambridge University Press,
England, 1973). See also \cite{review}  for a comprehensive discussion of
the $AdS$ spacetime. The Penrose diagram of the $AdS$ space with a flat 
domain wall was taken from \cite{hawking}.


\bibitem{gubser} S. S. Gubser, [hep-th/9912001].

\bibitem{hms} S. Hawking, J. Maldacena and A. Strominger, [hep-th/0002145].


\bibitem{no} S. Nojiri, S. D. Odintsov and S. Zerbini, Phys. Rev. D
(to be published) [hep-th/0001192]; S. Nojiri and S. D. Odintsov, 
Phys. Lett. B (to be published) [hep-th/0004097].

\bibitem{b-h} J. D. Brown and M. Henneaux, Comm. Math. Phys. {\bf 104},
207 (1986).


\bibitem{manifold-universe} N. Arkani-Hamed, S.
Dimopoulos, G. Dvali and N. Kaloper,
[hep-ph/9911386].

\bibitem{staro} A. A. Starobinsky, Phys. Lett. B {\bf 91}, 99 (1980).

\bibitem{gl} R. Gregory and R. Laflamme, Phys. Rev. Lett. {\bf 70}, 
2837 (1993).

\bibitem{gs} J. Garriga and M. Sasaki, [hep-th/9912118].

\bibitem{ca} J. D. Barrow, E. J. Copeland, E. W. Kolb and A. R. Liddle, 
Phys. Rev. D {\bf 43}, 984 (1991).

\bibitem{CKN} A. Chamblin, A. Karch and A. Nayeri, [hep-th/0007060]. 


\bibitem{matt} S. K. Blau, E. I. Guendelman and A. H. Guth,
Phys. Rev. D {\bf 35},
1747 (1987); M. Visser, Nucl. Phys. B {\bf 328}, 203 (1989).

\end{thebibliography}
\end{document}